\documentclass[twocolumn,showpacs,preprintnumbers,amsmath,amssymb]{revtex4}

\usepackage{graphicx}
\usepackage{bm}
\usepackage{amsmath}

\begin{document}

\title{Coherent Phonon Coupling to Individual Bloch States in Photoexcited Bismuth}

\author{E. Papalazarou$^{1}$, J. Faure$^{2,3}$, J. Mauchain$^1$, M. Marsi$^1$, A. Taleb-Ibrahimi$^4$, I. Reshetnyak$^2$, A. van Roekeghem$^2$, I. Timrov$^2$, N. Vast$^2$, B. Arnaud$^5$, L. Perfetti$^2$}

\affiliation{%
$^{1}$ Laboratoire de Physique des Solides, CNRS-UMR 8502, Universit\'e Paris-Sud, F-91405 Orsay, France}
\affiliation{%
$^{2}$ Laboratoire des Solides Irradi\'{e}s, Ecole polytechnique-CEA/DSM-CNRS
UMR 7642, F-91128 Palaiseau, France}
\affiliation{%
$^{3}$ Laboratoire d'Optique Appliqu\'ee, Ecole polytechnique-ENSTA-CNRS, F-91128 Palaiseau cedex, France}
\affiliation{%
$^{4}$ Synchrotron SOLEIL, Saint-Aubin-BP 48, F-91192 Gif sur Yvette, France}
\affiliation{%
$^{5}$ Institut de Physique de Rennes (IPR), UMR UR1-CNRS 6251, F-35042 Rennes Cedex, France}

\begin{abstract}

We investigate the temporal evolution of the electronic states at the bismuth (111) surface by means of time and angle resolved photoelectron spectroscopy. The binding energy of bulk-like bands oscillates with the frequency of the $A_{1g}$ phonon mode whereas surface states are insensitive to the coherent displacement of the lattice. A strong dependence of the oscillation amplitude on the electronic wavevector is correctly reproduced by \textit{ab initio} calculations of electron-phonon coupling. Besides these oscillations, all the electronic states also display a photoinduced shift towards higher binding energy whose dynamics follows the evolution of the electronic temperature.

\end{abstract}

\maketitle

It is well known that most elements crystallize in structures with hexagonal or cubic symmetry. One notable exception is bismuth, which instead crystallizes in the A7 rhombohedral structure. It has been originally proposed by Jones and Peierls that cubic bismuth would be unstable towards a lattice distortion leading to the doubling of the unit cell \cite{Peierls}. Due to the distorted nature of the groundstate, the atomic positions are very sensitive to the electronic distribution in the conduction band. This conjecture has been experimentally verified by suddenly changing the occupation number of the electrons via the absorption of a femtosecond laser pulse \cite{Dresselhaus,Merlin,Hase}. Due to photoexcitation, a large coherent phonon modulates the distance between the two atoms in the unit cell along the (111) direction. The displacement vector indicates that the lattice coordinates move towards a hypothetical phase with a single atom per unit cell. In reality the material displays a non-thermal melting before the symmetric phase can be attained \cite{Miller}.

The atomic motion following the photoexcitation of bismuth is nowadays well understood. Time resolved X-ray diffraction and Density Functional Theory (DFT) calculations could accurately describe the amplitude of the oscillations and the frequency softening at high excitation densities \cite{Fritz,Ingold,Arnaud}. On the other hand, the temporal evolution of electronic states has never been directly observed. Some aspects of the electronic dynamics can be inferred from general arguments. The generation mechanism of coherent phonons with large amplitude implies that low energy electronic excitations are coupled to the $A_{1g}$ mode. It is also expected that the electron-phonon coupling of such distorted material varies with the electronic wavevector. Nonetheless, the determination of the coupling matrix elements for individual Bloch states is still an experimental challenge. We approach this task by time resolved photoelectron spectroscopy measurements.
The electronic states of the bismuth (111) surface are characterized by angle resolved photoemission and first principles calculations. We confirm that the electronic structure displays a rich combination of bulk-like bands, surface states and surface resonances \cite{Hofmann,Kimura}. Upon photoexcitation, these electronic states display a dynamics that depends on wavevector and band index. The connection of our data with DFT calculations provides a full characterization of the coupling between electrons and the $A_{1g}$ mode. In addition, we observe that the electronic states do not follow rigidly the motion of the lattice displacement. Instead, the bands also display a purely electronic shift towards higher binding energy.

The experiments were performed with the FemtoARPES setup, using a Ti:Sapphire laser that generates 35 fs pulses centered at 810 nm with repetition rate of 250 kHz.
Part of the beam is employed to generate the fourth harmonic by frequency doubling in BBO crystals ($\beta$-BaB$_2$O$_4$). The 205 nm probe and the 810 nm pump are focused on the sample with a spot diameter of 100 $\mu$m and 200 $\mu$m, respectively. Their cross-correlation in a BBO crystal has a full width at half maximum (FWHM) of 80 fs. An electrostatic spectrometer with energy resolution better than 10 meV and angular resolution better than 0.5 degrees analyses the photoelectrons emitted by the 205 nm beam. The (111) surface of bismuth has been obtained by sputtering and annealing cycles of a single crystal. All measurements have been performed at the base temperature of 130 K and at the base pressure of 7$\cdot$10$^{-11}$ mbar. The out-of-equilibrium spectra have been collected with incident pump fluence of 0.6 mJ/cm$^2$.

\begin{figure} \begin{center}
\includegraphics[width=1\columnwidth]{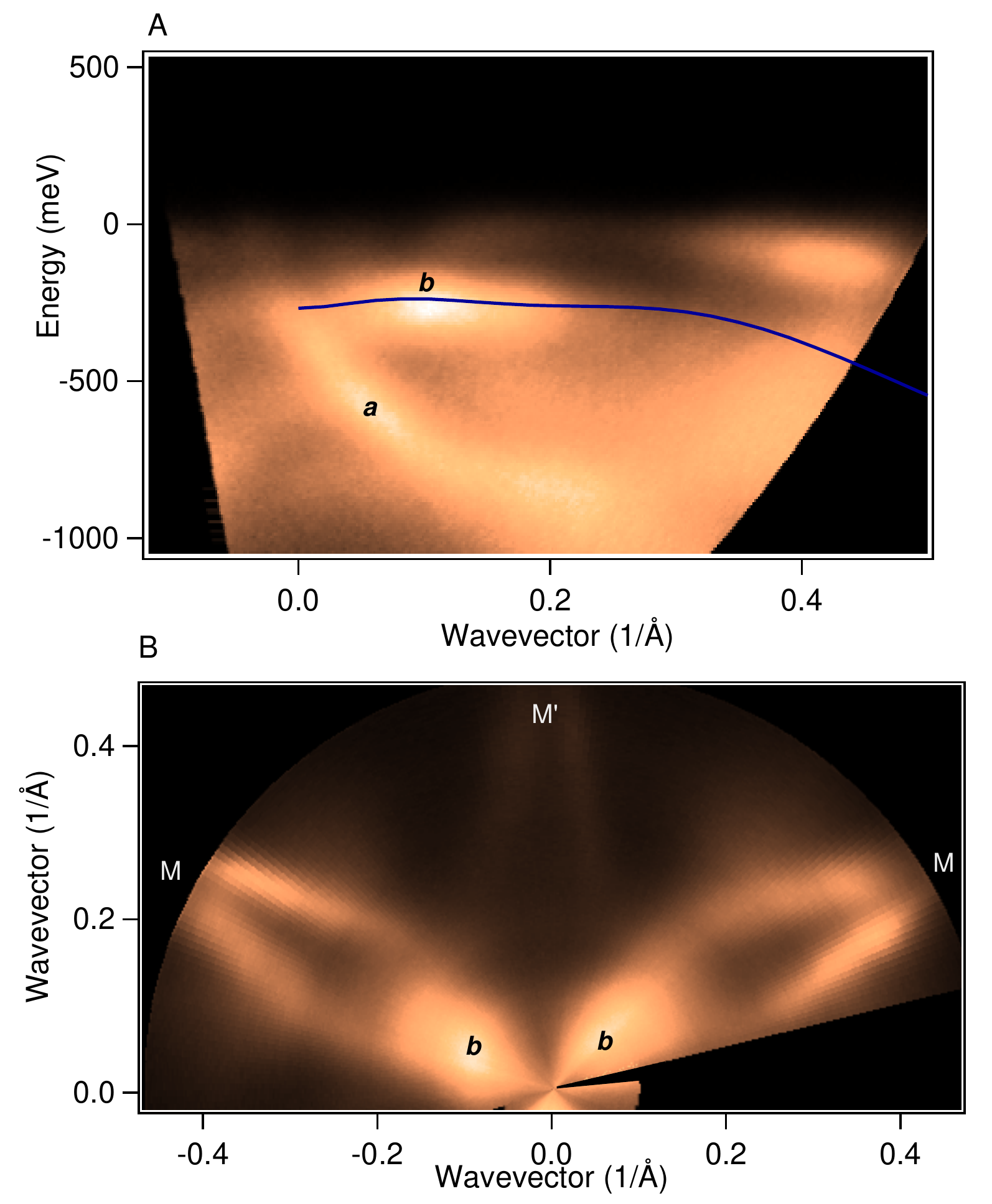}
\caption{A): Intensity map of photoelectrons emitted for parallel wavevector along the $\Gamma$-M direction. 
Band $a$ and band $b$ are the surface resonance and the bulk-like state, respectively. The blue line is a bulk band calculated for perpendicular wavevector $k_{\perp}=0.45$ \AA$^{-1}$. B): Intensity map of photoelectrons emitted at binding energy $\epsilon=-0.26$ eV. An intense structure originating from band $b$ is visible along the $\Gamma$-M direction but not along $\Gamma$-M'.} \label{Fig1}
\end{center}
\end{figure}

Figure \ref{Fig1}A shows an intensity map of the photoelectrons emitted along the $\Gamma$-M direction of the Brillouin zone. As previously reported, the break-down of translational symmetry in the (111) direction generates surface states that
intersect the Fermi level and give rise to a Fermi surface \cite{Hofmann}. These evanescent wavefunctions are localized at the topmost layers, thus conferring to the surface of bismuth good metallic properties. Although, the band structure supports surface states only near the Fermi level, some bulk-like states and surface resonances are clearly visible at higher binding energy \cite{Ast}. Two bands of different nature are indicated in Fig. \ref{Fig1}A by letter $a$ and $b$. Due to the matrix elements of the photoemission process, band $b$ is clearly visible only when the parallel component of the electronic wavevector is $k_{\|}<0.25$ \AA$^{-1}$. Band $a$ has larger binding energy, disperses more strongly and intersects band $b$ at the $\Gamma$ point.
The overall agreement between our measurements and one step photoemission calculations is  remarkable \cite{Kimura}. A. Kimura \textit{et al.} \cite{Kimura} also show that band $a$ has a large spin polarization whereas band $b$ does not. This finding suggests that $a$ is a surface resonance whereas $b$ is a bulk band.
In order to confirm it, we performed DFT calculations of bulk bismuth within the framework of the local density
approximation and the ABINIT \cite{Gonze} package. Since the photoemission process does not conserve the perpendicular component of the electronic wavevector, the connection between a peak in the photocurrent and a calculated band is formally not allowed \cite{Krasovskii}. Nonetheless, Fig. \ref{Fig1}A shows that the band calculated along the $\Gamma$-M direction for perpendicular wavevector $k_{\perp}=0.45$ \AA$^{-1}$ matches reasonably well the experimental intensity distribution.
More insights on the dispersion of band $b$ can be obtained by mapping in the reciprocal space the photoelectrons at fixed kinetic energy. Figure \ref{Fig1}B shows an intensity map of photoelectrons acquired at different wavevectors for binding energy $\epsilon=-260$ meV. The structure $b$ is very intense along $\Gamma$-M whereas it vanishes in the direction $\Gamma$-M'. 
The experimental evidence that band $b$ is invariant for a rotation of 120$^\circ$ but not for a rotation of 60$^\circ$ corroborates the bulk character of band $b$ \cite{Hofmann}.

\begin{figure} \begin{center}
\includegraphics[width=1\columnwidth]{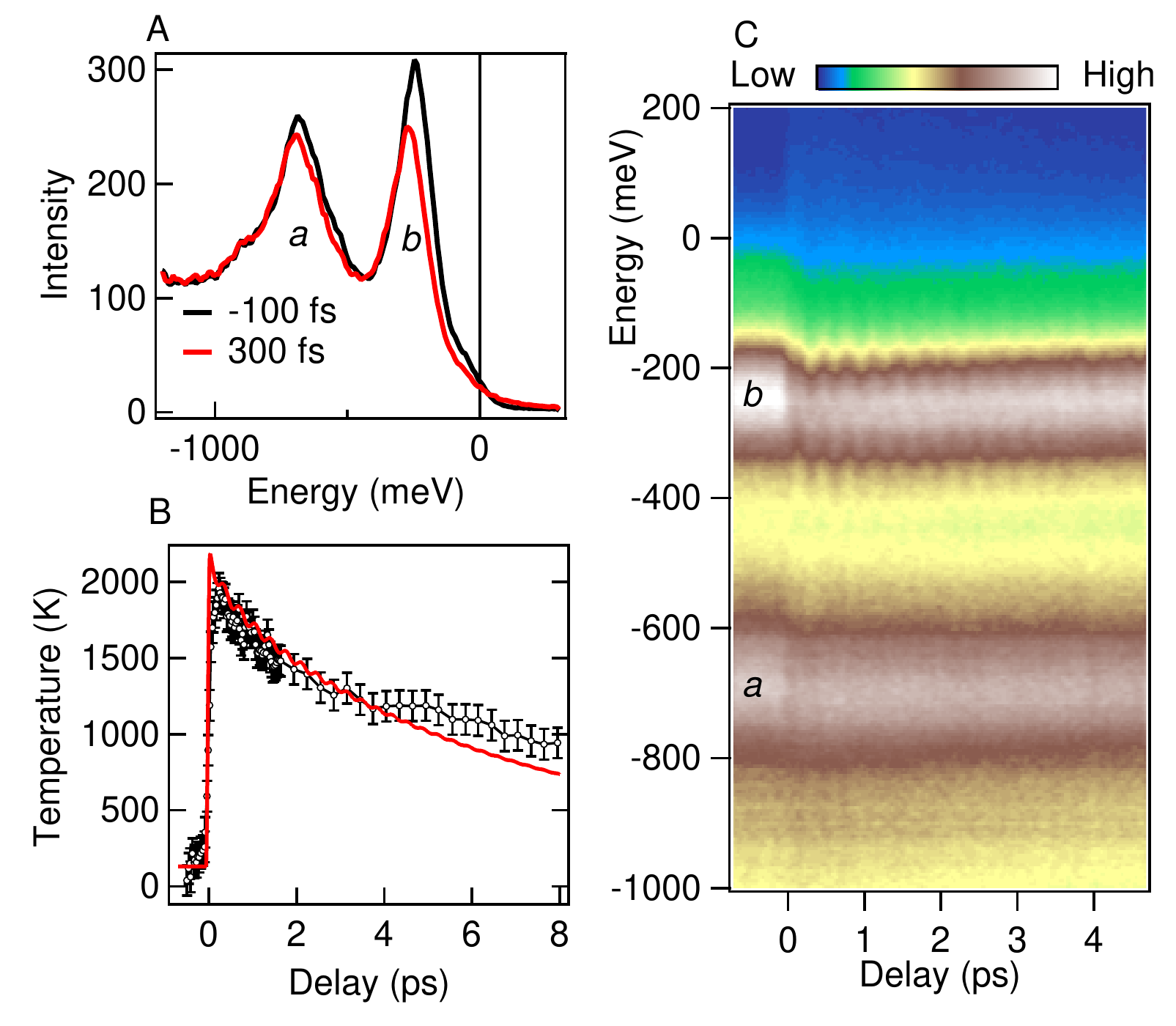}
\caption{A): Photoelectron spectra acquired 100 fs before and 300 fs after the arrival of the pump beam along the $\Gamma$-M direction at $k_{\|}=0.12$ \AA$^{-1}$. B): The effective temperature of photoexcited electrons as a function of pump probe delay extracted from the data (marks) and simulated with the model of ref. \cite{Arnaud} (red line). C): Intensity map of photoelectrons emitted at wavevector $k_{\|}=0.12$ \AA$^{-1}$ along the $\Gamma$-M direction as a function of pump probe delay.} \label{Fig2}
\end{center}
\end{figure}

Photoexcitation by an intense and ultrafast laser pulse generates sizable effects on the photoelectron current (see movie in the supplementary material). Figure \ref{Fig2}A shows the spectrum acquired at $k_{\|}=0.12$ \AA$^{-1}$ along the $\Gamma$-M direction for two different pump-probe delays. Note the large reduction of spectral weight taking place 300 fs after the arrival of the pump beam. The partial depletion of states $a$ and $b$ can be approximately described by a Fermi-Dirac distribution with an effective temperature of 1900 K. As shown by Fig. \ref{Fig2}B, the electronic temperature decays as a function of pump probe delay with a time constant of 6 ps. 
The relaxation of the electronic energy density at the surface is due to the combined effect of electronic diffusion and heat dissipation into the lattice bath. We could accurately reproduce the temporal evolution of the electronic temperature by solving the coupled differential equations in ref. \cite{Arnaud} 
with incident pump fluence of 0.6 mJ/cm$^2$,
electron thermal conductivity $k_0 = 0.04$ W m$^{-1}$K$^{-1}$ and a temperature dependent electron-phonon coupling.

Beside the large changes of the electronic distribution, we also observe strong variations in the structure of the electronic states.
As shown by Figure \ref{Fig2}A, both peak $a$ and $b$ shift towards higher binding energies after the arrival of the pump pulse. In addition, the position of peak $b$ displays periodic modulations at the frequency of the $A_{1g}$ mode. Such a dynamics arises from the coupling of the coherent phonon to electronic states and can be directly observed on the photoelectron intensity map of Fig. \ref{Fig2}C. The oscillations are large for the bulk band $b$ whereas they fall below the detection limit for the surface resonance $a$ and for all other surface states. This observation leads to two possible conclusions: either the amplitude of the $A_{1g}$ mode is smaller on the topmost bilayer or the surface states have a weak coupling to this mode.

\begin{figure} \begin{center}
\includegraphics[width=1\columnwidth]{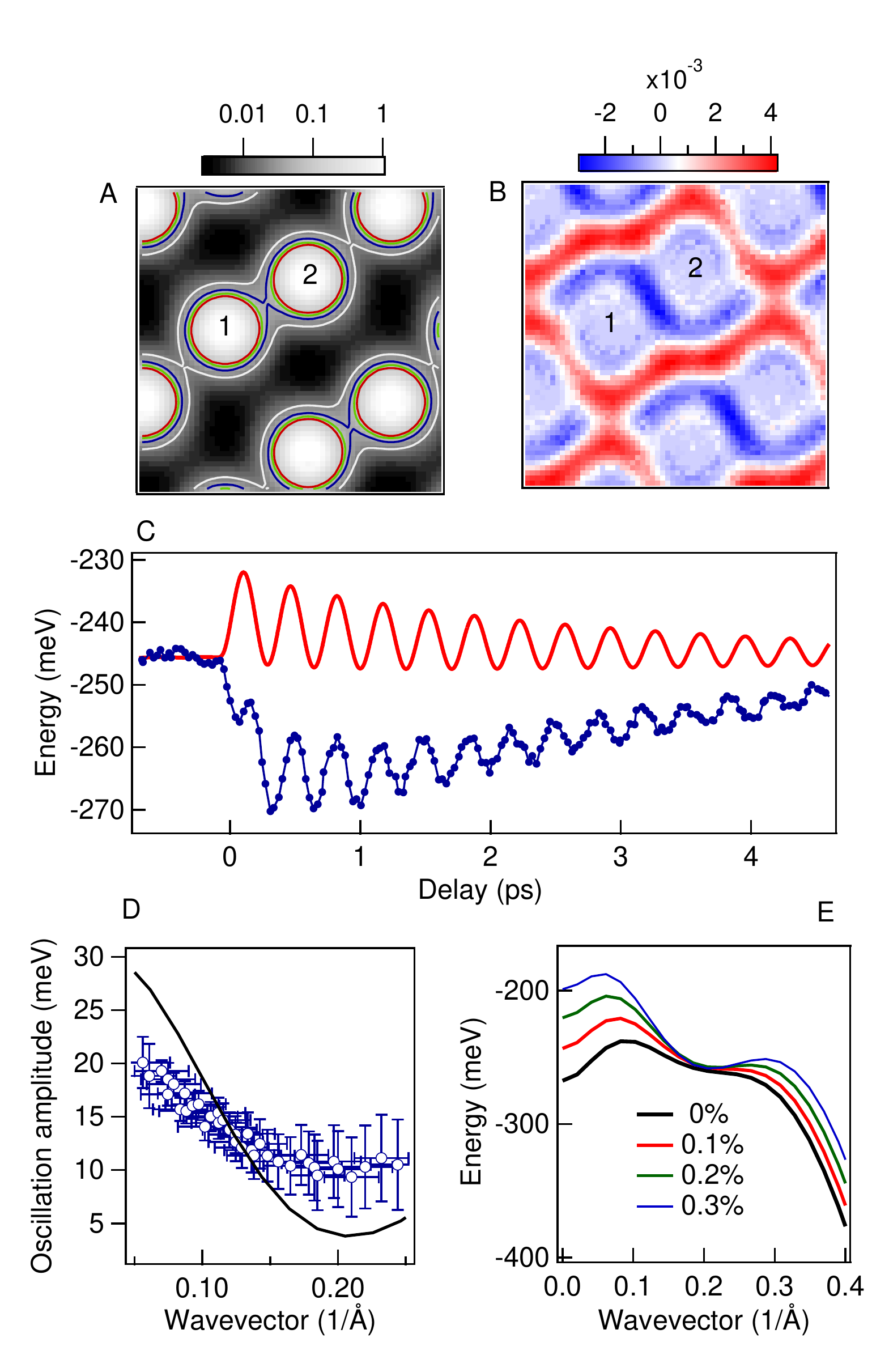}
\caption{A): Colorscale plot of the electronic density at 130 K in the (1$\bar{1}$0) plane. 
B): Relative change of the electronic density for  an increase of electronic temperature from 130 K to 2000 K. C): Binding energy of peak $b$ for $k_{\|}=0.12$ \AA$^{-1}$ along the $\Gamma$-M direction as a function of pump probe delay (blue marks) and temporal evolution obtained by \textit{ab initio} calculations (red line) \cite{Arnaud}. D): Oscillation amplitude of peak $b$ as a function of $k_{\|}$ (marks) together with the prediction of DFT theory for a lattice displacement $u=1.6$ pm (solid line). E): Calculated energy dispersion of band b along the $\Gamma$-M direction when the $A_{1g}$ phonon mode is activated ($ u \neq 0$). The calculations are performed for two atoms located at $\pm (0.2329\cdot c + u)$ along the diagonal of the unit cell for $u/c=0.1$ \%, 0.2 \%  and 0.3 \%.} \label{Fig3}
\end{center}
\end{figure}

The excitation of coherent phonons in semimetals has been the subject of many experimental and theoretical works \cite{Dresselhaus,Merlin,Hase,Miller,Fritz,Ingold,Arnaud}. Since bismuth absorbs photons at 810 nm, the generation mechanism is mainly due to a displacive excitation of the charge density \cite{Dresselhaus,Merlin}.
The DFT calculations can help us to illustrate this point: figure \ref{Fig3}A displays a colorscale plot of the electronic density  $\rho(r,T_i)$ at $T_i=130$ K in the (1$\bar{1}$0) plane. The sudden increase of electronic temperature from $130$ K to $T_f=2000$ K changes the spatial ditribution of electronic density. We show in Fig. \ref{Fig3}B that  $(\rho(r,T_f)-\rho(r,T_i))/\rho(r,T_i)$ is negative in the region between nearest neighbours (atoms 1 and 2). The transfer of electronic charge out of the bonding region increases the repulsion between the ion cores. As a consequence, the atoms oscillate around a new equilibrium position until the energy dissipation into other phonon modes brings the system back to the initial state. The lattice motion changes the binding energy of the Bloch state $| b,k_{\|}\rangle$ by $D_{b,k_{\|}} \cdot u$, where $u$ is the A$_{1g}$ phonon coordinate and $D_{b,k_{\|}}$ is the deformation potential. Figure \ref{Fig3}C shows the binding energy of peak $b$ at the base temperature of 130 K as a function of pump probe delay. The oscillations induced by the $A_{1g}$ mode have frequency $\nu=2.97 \pm 0.05$ THz and damping time $\gamma=2.6 \pm 0.2$ ps.
These values compares well with the frequency and damping time observed by transient reflectivity in comparable experimental conditions \cite{Hase}.
We reproduce the oscillations of peak $b$ by solving the model proposed in Ref. \cite{Arnaud} and using the value of the
computed deformation potential $D_{b,k_{\|}}=0.91$ eV/\AA~ for $k_{\|}=0.12$ \AA$^{-1}$. The simulated lattice dynamics attains a maximum displacement value $u=1.6$ pm, corresponding to 0.14 \% of the body diagonal length $c= 11.86$ \AA.


As shown in Fig. \ref{Fig3}D, the measured amplitude of oscillation is 20 meV for $k_{\|}<0.05$ \AA$^{-1}$ and decreases down to 10 meV for $k_{\|}<0.2$ \AA$^{-1}$. At larger wavevectors the photoelectron intensity of band $b$ is too low for a reliable extraction of the $D_{b,k_{\|}} \cdot u$ value.
The large $k_{\|}$-dependence of $D_{b,k_{\|}}$ has been previously reported for the amplitude mode of TbTe$_3$ \cite{Schmitt}. Due to the presence of a Charge Density Wave, the variation of the interaction strength as a function of the bloch state has been explained by the good nesting properties of the Fermi surface. On the other hand, our measurements suggest that the wavevector dependence of the electron-phonon coupling constitutes a general property of covalent crystals. This aspect of the coupling can be understood by a simple tight binding model on a linear chain: if the lattice displacement modifies the overlap between atomic orbitals the binding energy of the Bloch states will change differently for different wavevectors. Such $k$ dependence of the electron-phonon coupling is essential for several problems of solid state physics: it accounts for the presence of Charge Density Waves that do not respect the nesting condition \cite{Liu} and it allows for the correct estimate of the high transition temperature in superconducting MgB$_2$ \cite{Louie}. 

In the case of bismuth, our data offer an unique opportunity to test the deformation potential extracted by DFT. We calculate the band structure of bismuth for several values of the displacement $u$. 
Figure \ref{Fig3}E shows the dispersion of band $b$ for $u$ between 0.1\% and 0.3\% of the body diagonal length $c= 11.86$ \AA.  In agrement with the measurements, the binding energy of band $b$ strongly varies near the $\Gamma$ point whereas is less sensitive to the phonon coordinate for $k_{\|}=0.2$ \AA$^{-1}$. On the other hand, Figure \ref{Fig3}D shows that the wavevector dependence of $D_{b,k_{\|}}$ is stronger in the calculations than in the experiment. A more accurate simulation of the measured signal would require the computation of one-step photoemission spectra \cite{Kimura}.
 

\begin{figure} \begin{center}
\includegraphics[width=1\columnwidth]{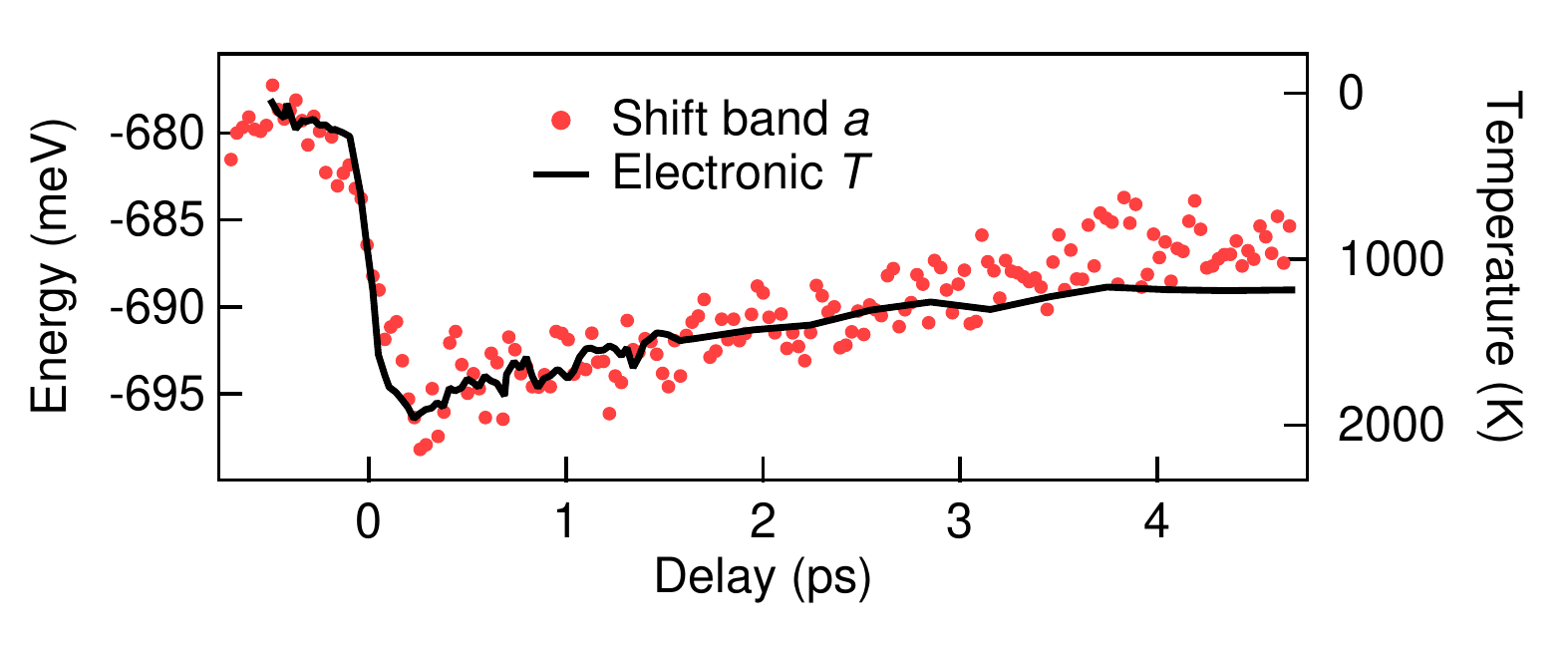}
\caption{ Peak position of the surface resonance $a$ for parallel wavevector $k_{\|}=0.12$ \AA$^{-1}$ as a function of pump probe delay (red marks, left axis), and electronic temperature (black line, right axis). 
} \label{Fig4}
\end{center}
\end{figure}

It is clear in Fig. \ref{Fig3}C that the dynamics of electronic states does not depend simply on the lattice coordinates. 
Besides coherent phonon oscillations, an additional effect shifts peak $b$ towards negative values. This photoinduced shift occurs also in the surface resonance and surface states. As shown in Fig. \ref{Fig4}, the dynamics of peak $a$ nearly follows the electronic temperature. Under similar experimental conditions, the photoinduced increase of binding energy as been observed also in the surface states of Gadolinium \cite{Bovensiepen}. An opposite behaviour has instead been reported in the electronic states of materials with photoinduced phase transitions \cite{Schmitt,Perfetti}. 

The detailed analysis of the electronic shift as a function of $k$ will be part of fore coming work. Nonetheless, we notice that the magnitude of this component depends on the electronic wavevector and band index. 
A wavevector dependent stiffening of the electronic states in the bonding band can be expected because of the reduced ion core screening at elevated electronic temperature. However, the finite temperature DFT calculations indicate that the charge transfer of figure \ref{Fig3}B accounts at maximum for 2 meV to the shift observed in band $b$. This discrepancy may be due to the fact that our bulk model of the electronic states does not take into account charge redistribution and carrier transport at the surface of the sample. Alternatively, the dynamical aspect of the electronic screening may require simulations that go behind the DFT method \cite{Faleev}.

In conclusion, the electronic states of photoexcited bismuth oscillate with the frequency of the A$_{1g}$ mode. By analyzing the photoelectron spectra collected at different angles, we reconstruct the deformation potential of specific Bloch states. A quantitative agreement with DFT calculations confirms that the photoinduced charge transfer accounts for the amplitude of the observed oscillations.
At last, we mention that a purely electronic effect adds to the temporal evolution of the electronic states with the dynamics of the electronic temperature.

The FemtoARPES project was financially supported by the RTRA Triangle
de la Physique, the ANR program Chaires d'Excellence (Nr. ANR-08-CEXCEC8-011-01). We also aknowledge the contributions of DGA and PNANO ACCATTONE. Part of the Density Functional Theory results have been obtained  or reproduced also by the Quantum Espresso \cite{Giannozzi:2009} package.
Calculations were performed using HPC resources from GENCI-CINES (Nr. 2010-09596 and project 2210).

\newpage

\end{document}